\documentclass{nature}
\usepackage{subfig}
\usepackage{enumerate}
\usepackage{graphicx}
\graphicspath{{pics/}}
\usepackage{amsmath}
\usepackage{amsthm}
\usepackage{amsfonts}
\usepackage{float}

\theoremstyle{plain}
\newtheorem{thm}{Theorem}[section] 
\newtheorem{lem}{Lemma}[section] 
\theoremstyle{definition}

\newtheorem{proof}{Proof}[section]

\title{Three Dimensional Aggregation of Magnetic Particles}

\author{Alex Pai$^1$, Dimitar Ho$^2$ \& Ali Hajimiri$^1$}

\begin{document}

\maketitle

\begin{affiliations}
 \item Department
of Electrical and Medical Eng\-ineering, Cali\-fornia Inst\-itute of Tech\-nology, Pasadena,
CA, 91125 USA
 \item Department of Control and Dynamical Systems,Cali\-fornia Inst\-itute of Tech\-nology, Pasadena,
CA, 91125 USA
\end{affiliations}
\begin{abstract}
Magnetic drug delivery is a promising therapeutic because of magnetic fields' ability to permeate unperturbed in human tissue. One of the long-standing challenges in magnetic drug delivery is the inability to generate 3D aggregation non-invasively within the interior of the body. Earnshaw's theorem, which proves the impossibility of creating an energetic minimum in a curl-free and divergence-free field such as a magnetic field. However, one of the assumptions of Earnshaw's theorem is a \textit{static} field. Here we show that it is possible to utilize a \textit{dynamically} changing field and a dissipative force such as the drag, which is generally present, to create a stable aggregation point for magnetic particles. We also introduce a theoretical framework for designing the suitable magnetic fields for controlling a given magnetic particle in a particular fluid. This framework enables accurate determination of the necessary parameters for aggregation across a wide variety of magnetic particles and across multiple biologically-relevant fluids. By coating magnetic particles with desired therapeutic agents or attaching them to cells, a new class of treatment methodologies can be created in therapies such as targeted drug delivery and cell-based therapies. By dynamically changing the aggregation point, agents can also be guided along a particular path in the body. This technique of using dissipative forces to create a stable 3D aggregation point for particles could possibly be extended to a broad range of applications such as microscopic and macroscopic manipulation, robotics, guided self-assembly, magnetic plasma confinement, tissue engineering, and ion traps for quantum computers.
\end{abstract}

\begin{figure}[h]
\centering
\includegraphics[height=2.5in]{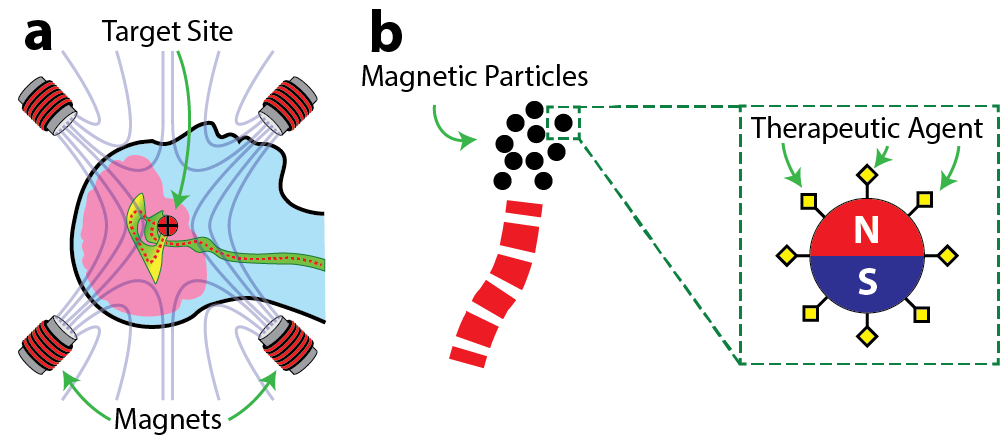}
\caption{\textbf{Magnetic Localization} \textbf{a}, Magnetic localization can be used to guide magnetic particles throughout the body using magnets located outside the body. One possible application is guiding particles through the ventricular system to a tumor site. \textbf{b}, Magnetic particles can be coated with a specific therapeutic agent specific for treating a disease.}
\label{system}
\end{figure}

Magnetic targeting is a promising technology for treatment of a wide array of diseases \cite{chomoucka2010magnetic, chorny2011magnetic, chertok2008iron, veiseh2010design, white2015functionalized,amirfazli2007nanomedicine}. Paramagnetic iron-oxide nanoparticles (IONPs) are coated with therapeutic agents or attached to cells, and then focused to a target site in the body (Fig. \ref{system})\cite{TIETZE2015463}. This allows for effective treatment by localizing therapy only to affected areas while minimizing the undesirable side-effects of healthy cells. IONPs are typically superparamagnetic (only magnetized when exposed to a magnetic field \cite{}) and generally have little or no toxicity \cite{tassa2011dextran}. IONPs have commonly been used to label and track cells \textit{in vivo} with magnetic resonance imaging (MRI) \cite{loebinger2009magnetic, thu2009iron}. Recently, IONPs have also been used for cell localization using magnetic fields \cite{polyak2008high, pislaru2006magnetically, nishida2006magnetic, kyrtatos2009magnetic}. In these applications, the magnetic fields were derived from either implanted magnetized objects or placing large conventional magnets close to the skin. Other potential techniques use a combination of continuous tracking and field adjustment to traverse towards a target site\cite{doi:10.1002/wnan.1311}. 

Traditionally, 3D aggregation of magnetic particles has been achieved using magnetic fields and feedback control \cite{}. 
Feedback control is undesirable because it requires high temporal and spatial resolution sensing or imaging of the particles. Even assuming the availability of sufficient feedback information, designing a feedback controller to stably aggregate multiple particles remains a difficult problem because the controller can only control particles in an ensemble rather than each particle individually\cite{doi:10.1002/wnan.1311}. Nevertheless, feedback control has been deemed necessary because of the supposed impossibility in creating an energetic minimum between permanent and/or electromagnets (except under impractical conditions such as diamagnetism). Although Earnshaw's theorem has been used as justification of this impossibility, the underlying assumption relies on a static magnetic field and frictionless environment\cite{nacev2012towards}. We will show in this paper that it is possible to stably aggregate magnetic particles without feedback in the presence of sufficient drag. This approach can generate a 3-dimensional stable aggregation point such that a multitude of magnetic particles will be simultaneously focused to that point. The aggregation point position can also be adjusted to move particles along a path. This has significant implications in fields such as magnetic drug targeting. This aggregation method is a complete feed-forward system and thus does not require imaging.

Routing paramagnetic particles to a specific site requires the ability to generate a magnetic field profile that produces the necessary magnetic field intensity and gradient (and thereby force), that can be controlled for location, direction, and strength. In many practical applications, target sites can be located deeply beyond accessible surfaces (e.g., the skin). Use of a magnetic field is potentially superior to many other particle routing and targeting modalities due to its deeper depth of penetration and no known ionization effect. Additionally, the distribution and thus effectiveness of paramagnetic particles within the body can be readily evaluated using MRI as secondary verification. Although simple permanent magnets have been used to attract paramagnetic particles to the sites of interest in animals, the lack of control over the field profile makes it difficult to obtain the desired level and direction of force necessary to route and retain the intended magnetic particles at the target site in three dimensions. Furthermore, this method can only localize particles to the surface of the body. 

To target locations inside the surface of the body, attraction schemes utilize external magnetic manipulation and imaging to guide a particle directly to a target area \cite{doi:10.1177/0278364908100924}. However, implementing these mechanisms requires constant imaging feedback, which makes it unfeasible for long-term applications. A sequence of unstable fields must continuously be modified depending on the trajectory of particles while simultaneously imaging them. Implanted magnets are invasive to the body and their position is not easily adjusted \cite{doi:10.1118/1.4805097}. Other schemes limit the need for imaging by passing magnetic particles back and forth across a target volume to increase the transit time in that volume \cite{6200498}. However, these come at a trade-off of specificity and increased treatment time. 

One of the critical limits of these techniques relies on the inability to create a 3D magnetic stability point (See Supplementary Discussion \ref{pseudo}). This limitation is governed by Earnshaw`s theorem \cite{earnshaw}.  Some demonstrations of pseudo-stability rely upon external forces (e.g., gravity), effective negative susceptibility (using weak diamagnetic fields or moving in a high susceptibility volume), or restricting dimensions to 1D or 2D for a saddle point in 3D to appear as a stable point in lower dimensions. These methods are not feasible for targeting inside the body. Another promising magnetic targeting method utilizes MR imagers to generate pulsed magnetic fields to a target. This has the advantage of extremely high field strengths and the ability to simultaneously image the target using MRI. However, permanent magnet based methods are able to generate higher gradient fields. Additionally, fields are still subject to Earnshaw's theorem and therefore can only bias magnetic particles towards one end of a sample. Targeting with increased resolution in the body requires a more precise method.

We propose a feedback-free scheme that breaks the assumption of Earnshaw's theorem to create a 3D stable aggregation point in a volume free of permanent magnets or electrical current. Our method for the stable attraction of paramagnetic particles utilizes a dynamically changing magnetic field to bypass the assumptions of Earnshaw's theorem and achieve stable 3D particle control without feedback. The magnetic field can be generated by a collection of independent electromagnets acting in concert. These elements act in concert to create a temporally varying field necessary for magnetic targeting. Alternatively, the field can be created with a mechanically moving permanent magnet. We show this phenomena experimentally with paramagetic particles in the presence of moving rare earth magnets.

\begin{figure}[H]
\centering
\includegraphics[width=6in]{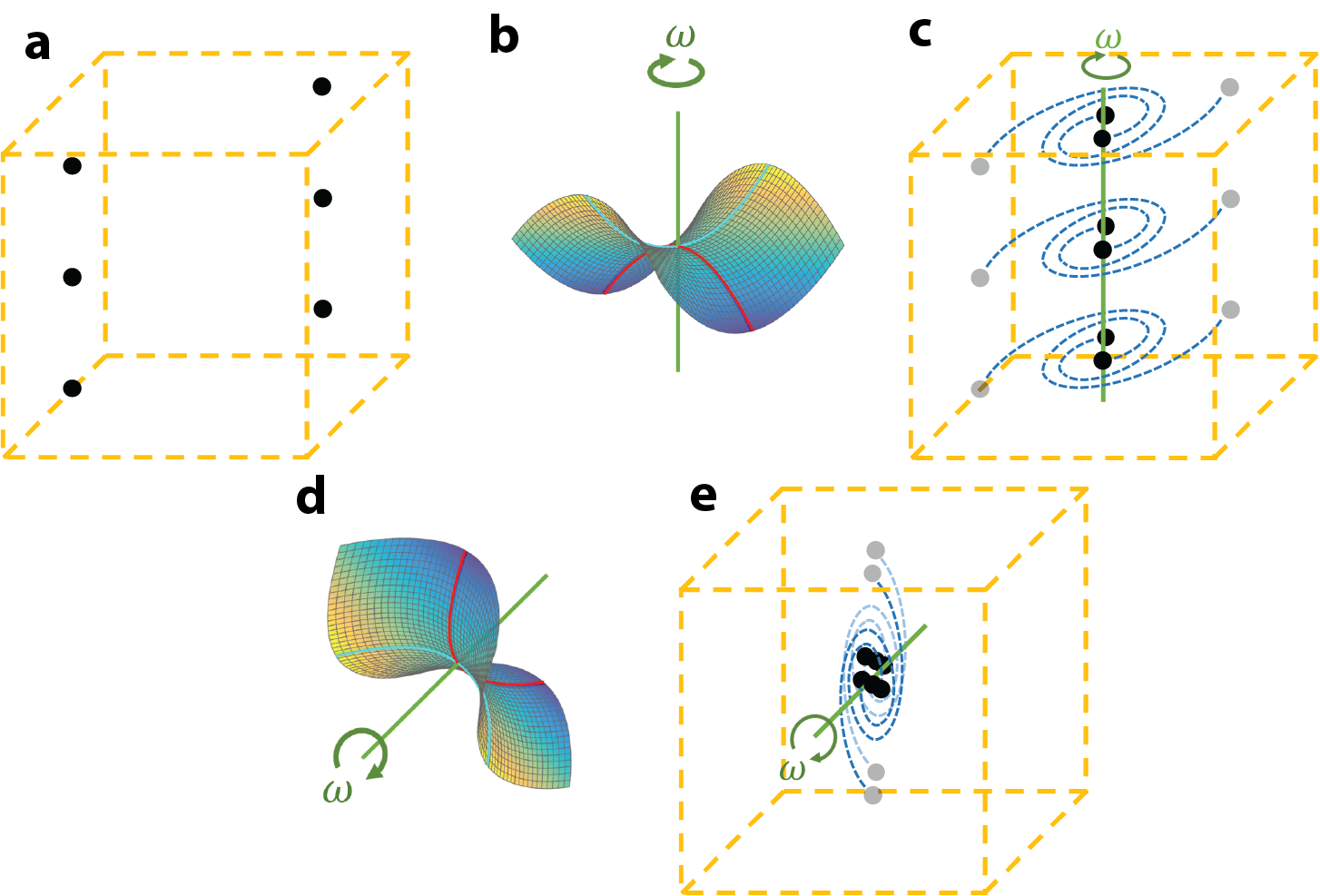}
\caption{\textbf{Method of 3D aggregation in a volume} \textbf{a}, Particles distributed throughout the volume of space are aggregated to a point inside the volume. No permanent magnets nor electromagnets need to be located within the volume. Additionally, a imaging/feedback mechanism is not necessary for localization of particles. The center of the volume is chosen for simplicity, but any point in the volume can be chosen. \textbf{b}, Particles are subject to a rotational magnetic force field. The frequency of rotation depends on the magnetic field, particle mass, and drag properties. \textbf{c} During the first rotation, particles are aggregated into a column. \textbf{d}, A second rotational magnetic field rotates along an axis orthogonal to the first rotational magnetic field. \textbf{e}, During the second rotation along a different axis, particles are aggregated from a column to a point.}
\label{fig_spiral}
\end{figure}

Our strategy for 3D aggregation is a three step process (Fig. \ref{fig_spiral}). First, we identify a set of fields that aggregates in two dimensions (Fig. \ref{fig_spiral}\textbf{b}). This set of fields is a rotational magnetic force field. At any instant in time, the fields are governed by Earnshaw's theorem. The rotational frequency and field composition should be chosen depending on the properties of the system (particle mass and drag). The rotational force field utilizes the drag of the particles to achieve stability. Second, we apply this field to the particles distributed in 3D (Fig. \ref{fig_spiral}\textbf{a}), aggregating them to a column of particles (Fig. \ref{fig_spiral}\textbf{c}). The column corresponds to the axis of rotation. Third, another unique axis of rotation should be chosen to aggregate the column of particles to a point (Fig. \ref{fig_spiral}\textbf{d}). The second and third steps can be repeated to localize particles within desired bounds (Fig. \ref{fig_spiral}\textbf{e}). To formalize this strategy, we first discuss the theory of magnetic force fields, then the criterion for stability given a rotational force field. We will also introduce the Simplified Linear Model (SLM) for describing the motion of the particle undergoing a drag force that scales linearly with velocity and magnetic force that is linear with position. Although this model describes stability at the origin, it is valid for any position using a change of variables. We use this model to estimate the criterion for stability and use them to implement a magnetically stabilized system.
We can calculate the stability of a magnetic particle in a 2D rotating magnetic force field using the particle and fluid properties.
The mass, \textit{m}, of a isotropic particle is given by its radius, \textit{r}, and density $\rho$, using the equation,
\begin{equation}
m= \frac{4}{3} \pi \rho r^3
\end{equation}
For a particle traveling in $\mathbf{x} \in \mathbb{R}^2$ undergoing Stokes's drag, the drag force, $\mathbf{F_d}$, is linearly related to the particle's velocity, $\dot{\mathbf{x}}$, and is given by,
\begin{equation}
\mathbf{F_d}=-b\mathbf{x}
\end{equation}
with the drag coefficient calculated as
\begin{equation}
b = 6 \pi \eta r
\end{equation}
where $\eta$ is the dynamic viscosity of the surrounding fluid. 
The motion of a paramagnetic particle undergoing Stokes's drag with drag coefficient $\textit{b}$ and experiencing a rotating magnetic field $\mathbf{H}(\mathbf{x},\omega t)$ at frequency $\omega$ is modeled via the following ordinary differential equation:
\begin{equation}\label{eq:sys}
 \ddot{\mathbf{x}} + 2\epsilon \dot{\mathbf{x}} = \alpha \nabla U_H\left(\mathbf{x},\omega t \right), \quad U_H = \left\|\mathbf{H}\left( \mathbf{x},\omega t\right) \right\|^2_2
\end{equation}
The constants $\epsilon$ captures the amount of damping in the dynamics due to friction, while $\alpha$ describes the ratio between the strength of magnetic and inertial forces present in the system. These constants are computed as
\begin{align}
\epsilon = \frac{b}{2m} = \frac{9 \eta}{4\rho r^2}, \quad \alpha = \frac{\mu_0 \chi}{2 \rho} 
\end{align} 
For a spherical particle with intrinsic magnetic susceptibility $\chi_i$, the effective volume susceptibility $\chi$ is calculated as:
\begin{equation}
 \chi = \frac{3\chi_i}{3+\chi_i}.
\end{equation}
The term $U_H$ plays the role of a magnetic potential in the dynamics, that only depends on the magnetic field.
\noindent We seek to construct a magnetic field that approximates a saddle potential at the origin. Mathematically, this translates to the condition $\nabla \mathbf{H}(0) \approx 0$. In Supplementary Equations \ref{seq}, we show that if we rotate this saddle potential at the right frequency, we can use the dissipative forces to stabilize the system. Moreover, to enable this stabilizing effect, we define the geometric quantities $c_{\delta}$ and $c_{\mu}$ such that
\begin{align}
c_{\mu} &= \frac{1}{2}\left.\nabla^2 U_H\right|_0\\
c_{\delta} &= \sqrt{\frac{1}{4}\left( \left.\nabla^2 U_H\right|_0\right)^2 - \mathrm{det}\left(\left. \mathcal{H}\left( U_H \right)\right|_0\right)}
\end{align}
where $\mathcal{H}$ denotes the Hessian operator.
These quantities describe the curvature of the saddle at the origin. 
\noindent More precisely, if the condition 
\begin{align}\label{eq:cond_r0_gen}
c_{\delta} > 2\sqrt{c_{\mu}}\sqrt{\frac{\epsilon^2}{\alpha }+c_{\mu}}
\end{align}
is satisfied, then the solutions of the ordinary differential equation \eqref{eq:sys} converge exponentially i.e. $\left\|x(t)\right\|_2 \sim \exp\left(-\lambda_x t\right)$ for some $\lambda_x >0$, if the frequency of rotation is chosen to lie within the range
 \begin{align}
\sqrt{\sqrt{\alpha^2 c^2_{\delta}+4\epsilon^2\left(\epsilon^2+\alpha c_{\mu} \right)}-2\epsilon^2-\alpha c_{\mu}} < \omega < \sqrt{\alpha\frac{c^2_{\delta}}{4c_{\mu}} - \epsilon^2}
\end{align}
As illustrated in (Fig. \ref{fig_spiral}), to achieve localization in 3D, our strategy is to alternate between the two axis of rotation (Fig. \ref{fig_spiral}\textbf{b} and Fig. \ref{fig_spiral}\textbf{d}) in a fixed time interval $T$. As presented in the control theoretic derivation in the Supplementary Equations \ref{seq}, we can show that we can always pick a long enough time-interval $T$ such that the particle converges exponentially fast to the origin in 3D even in the presence of slow unstable drift dynamics. This is a sufficient condition for stability. Practically speaking, $T$ must also be less than maximum available stability time to allow for at least two rotations.
\begin{figure}[h]
\centering
\includegraphics[width=\textwidth]{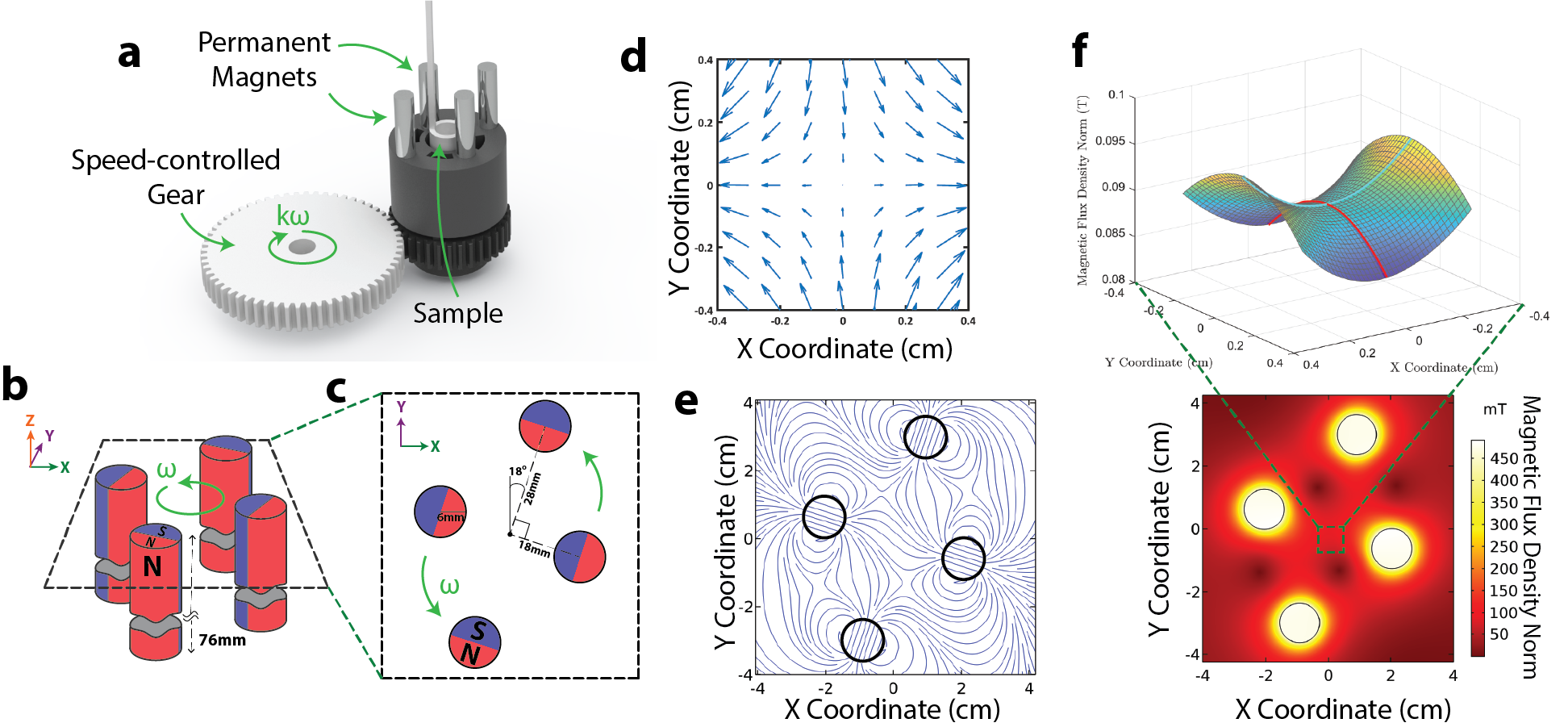}
\caption{\textbf{Experimental Setup Design.} \textbf{a}, The rotational magnetic field is created using a set of four diametrically magnetized rare earth magnets surrounding a sample. The magnets' rotational speed ($\omega$) is controlled by a separate speed-controlled gear rotating at $k\omega$ where $k$ is the gear ratio between the two gears. \textbf{b}, Isometric View of Magnets. \textbf{c}, Top View of Magnets. The magnets are arranged in a parallelogram. \textbf{d}, Magnetic Force Vector Field. \textbf{e}, Magnetic Field Lines of Setup. \textbf{f}, Magnetic Flux Density Norm of Setup.}
\label{experimentalSetup}
\end{figure}

The experimental setup used to demonstrate our stabilizing mechanism is shown in Fig.\ref{experimentalSetup}. A rotational magnetic potential approximating a saddle surface is used to localize particles to the axis of rotation. A variable speed motor is used to rotate a set of permanent magnets (Fig. \ref{experimentalSetup}\textbf{a}). A dish of the fluid sample containing a magnetic particle is fixed in the center of the rotating magnets. The orientation of the magnets are shown in Fig. \ref{experimentalSetup}\textbf{b,c}. Due to the vertical symmetry of the magnets, the magnetic force in the vertical dimension is virtually zero. The magnets generate a force field on magnetic particles. The vector field of force is shown in Fig. \ref{experimentalSetup}\textbf{d}. The force on the particles are strictly a function of position. There exists an approximately linear relationship between the Cartesian coordinates of the position and the magnetic force vector on a particle at that position. The magnetic field lines used to generate the force are shown in Fig. \ref{experimentalSetup}\textbf{e}. From the magnetic flux density norm, we can see that the magnetic potential forms a saddle surface around the stability point (Fig. \ref{experimentalSetup}\textbf{f}). The saddle surface rotates about the stability point with frequency $\omega$.

\begin{figure}[h]
\centering
\includegraphics[height=2.9in]{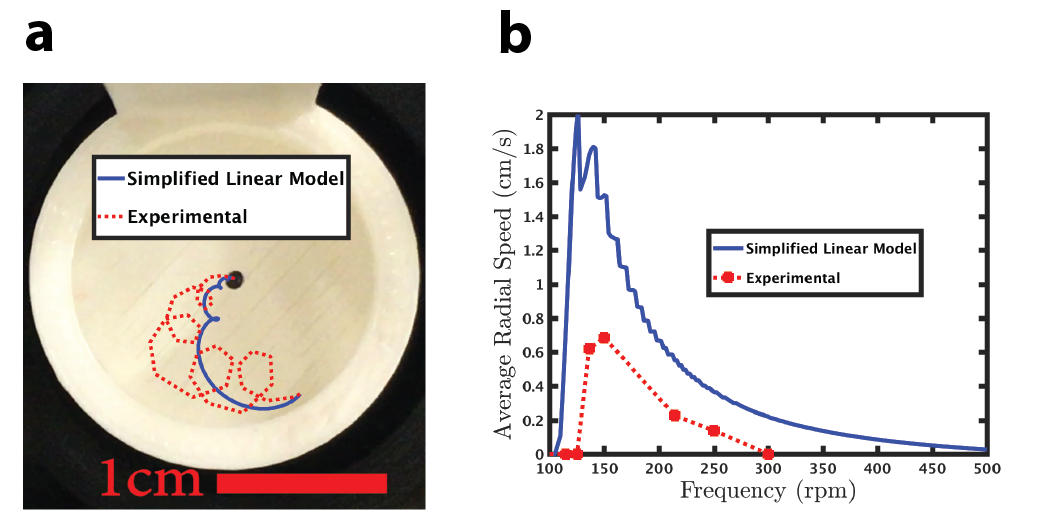}
\caption{\textbf{Aggregation Speed} \textbf{a}, Magnetic Particle Trajectory. The particle is a 1.09mm diameter magnetic particle suspended in water. The magnetic rotational frequency in this example is 150rpm. \textbf{b}, Average Localization Speed. A high speed indicates faster aggregation towards the target site (center of dish).}
\label{graphs}
\end{figure}

A comparison between trajectories simulated using our SLM and tracked experimental trajectories is shown in Fig. \ref{graphs}\textbf{a}. The average radial speed of the magnetic particle for various rotational frequencies are shown in Fig. \ref{graphs}\textbf{b}. Simulated trajectories using the SLM model overestimate radial speed of the particle, resulting in a larger anticipated frequency window for particle aggregation. There are a few factors that may contribute to these discrepancies. The dynamical model used for simulating trajectory calculates the translational motion of the particle using the state variables of the position and velocity of the particle. However, the model does not consider the rotational dynamics of the particle. In addition, our model assumes Stokes's drag with laminar flow. This simplifying assumption is more accurate for low Reynold's numbers. The order of our Reynold's number for this system is  \textasciitilde10. This implies flow near the boundary of laminar and turbulent flow. The SLM becomes a more accurate model with a smaller Reynold's number. Therefore, it more accurately describes the drag force for lower particle velocities, smaller diameter particles, less dense fluids, and higher dynamic viscosities. Additional sources of error that contribute to model inaccuracy include the nonlinearities in magnetic force field and fluid/air interface effects on particle drag.

In conclusion, we have demonstrated a theoretical framework of aggregating paramagnetic particles in 3D. Control variables should be chosen depending on system parameters and objective (maximum stability margin, aggregation speed, or oscillation minimization). The system is attractive to biological applications because the body has a very small magnetic response. Additionally, the distribution of paramagnetic particles can be readily imaged, but this technique does not rely on constant imaging feedback for particles to be guided to a target site. Localization occurs independent of initial position or motion. To verify this framework, we use the finite-difference approximation method to evaluate the motion of particles. This simulation was consistent with the theoretical framework. Additionally, we utilized an experimental setup to validate the principles of our localization scheme. We demonstrated aggregation using a rotational magnetic field, one of the core components to 3D targeted aggregation.

Although this work focuses specifically on the localization of mm-sized magnetic particles, this scheme can be extended to other size ranges as well. Our method stabilizes particles undergoing a force field in the presence of drag. We analyze stability under the assumptions of a homogeneous fluid of constant viscosity and minimal flow. This could limit the applicability of this method for biological and medical applications, where physiological structures such as the extracellular matrix and fluid flow such as hemodynamics can complicate aggregation. Although this limitation warrants further study, a potential work-around includes the injection of saline solution and nanoparticles around target areas such as organs. This allows the nanoparticles to travel freely inside the fluid cavity and access regions on the periphery of the cavity and surface of the organ.  Another work-around is the creation of vesicles of a nanoparticle/fluid solution. Particles inside the vesicle would be biased towards one direction and push against the vesicle wall. This would allow the vesicle to travel in regions without ideal bulk fluid properties. The vesicle wall could be coated with chemotherapy. Two forces are necessary for aggregation of particles: a guiding force and a dissipative force. In our example, we use magnetic force and drag force. The magnetic force acting on particles can also be substituted with other force fields such as electric or gravitational. Drag force can be substituted with other dissipative forces as well. Consequently, this is applicable to other systems. Earnshaw's theorem was first introduced for electrostatic configurations. Thus, this localization scheme can be extended for stabilizing charged particles in a dynamically changing electric field. Additionally, this method can be applied to localization of plasma in magnetic confinement fusion.

\begin{methods}
\subsection{Rotational Magnetic Field}

A sufficient condition for 3D aggregation is a properly configured pair of rotating magnetic fields about two distinct axes. Thus, a rotational magnetic field is one of the core components of our localization strategy. This can be generated in one of two ways. The first is through stationary electromagnets. The strength and polarity of each electromagnet can be dynamically changed to approximate a rotational magnetic field. The second method is by mechanically rotating a set of permanent magnets. In this work, the latter method was chosen.  To experimentally show the stabilizing phenomena, a rotational magnetic field is created that aggregates paramagnetic particles in two axes and has no effect in the third axis. The experimental setup is shown in Fig. \ref{experimentalSetup}\textbf{a}. The four permanent magnets used to generate the magnetic field are N48 Neodymium rare earth diametrically magnetized cylinders of 13 mm in diameter x 76 mm in length (CMS Magnetics). They are arranged in a parallelogram as shown in Fig. \ref{experimentalSetup}\textbf{b,c}. The vertical length of the magnet is long compared to horizontal dimensions to ensure no significant magnetic force on the particles in the vertical dimension. All components are constructed of non-magnetic materials and the motor is placed away from the sample to prevent magnetic interference. The magnets were placed within a plastic enclosure. The bottom of the enclosure is mounted onto a plastic gear. The gear is attached to a non-magnetic aluminum shaft, allowing for only rotational movement. The speed of rotation is controlled via a motor-mounted plastic gear. This is advantageous to mounting the enclosure directly to the motor because it isolates the iron shaft and magnetic fields generated by the servo from the paramagnetic particle. The motor (High Torque Nema 23 CNC Stepper Motor) is driven using a Nema TB6600 Driver and Arduino Uno. A plastic dish used to contain the fluid-particle sample and is suspended from above to isolate the sample from motor vibrations. All plastic components were produced using polylactic acid filament in a 3D printer (Makergear M2). Design files are available for 3D printed parts.
In Fig. \ref{experimentalSetup}\textbf{d}, we see that the experimental setup creates a nearly linear magnetic force field, where the magnetic force acting on the particles is a function of the position of the particle. The magnetic field lines can be seen in Fig. \ref{experimentalSetup}\textbf{e}. The square of the magnetic flux density norm (Fig. \ref{experimentalSetup}\textbf{f}) is linearly related to the potential energy and gives a sense of the energetics of the system. A energetic saddle point is created in the center of the setup. Despite the lack of an energetic minimum, a stability point can still be created using this saddle point.
\subsection{Susceptibility Measurements}

The paramagnetic particles are BKMS-1.2 polyethyene microspheres with a density of 1.2 g/cm$^3$ and a mean diameter of 1.09 mm (Cospheric). The susceptibility of the paramagnetic particles was measured using a MFK1-FA Kappabridge susceptometer (Agico). The magnetic volume susceptibility of the particles at room temperature (25$^{\circ}$ C) was measured to be 1.0036. 
The dynamic viscosity of water at room temperature is 8.90 x 10$^{-4}$ Pa*s. 
\subsection{Particle Tracking}

Particle movement are captured on video at a frame rate of 30 fps. Particle trajectories are determined in post-processing using the TrackMate plugin of image tracking software Fiji. To calculate the average radial aggregation speed shown in Fig. \ref{graphs}\textbf{b}, the total distance traveled from the starting position of the particle to the center of magnetic rotation was divided by the total time of travel. The starting position of the particle was always located on the periphery of the water dish. Thus, the total distance traveled for all experiments was 0.9 cm. The total time of travel is measured between the start of magnetic rotation and the time when the particle reaches is within one radial length (0.5 mm) of the destination. 

\subsection{Particle Motion Simulation}
The magnetic fields of the experimental setup are simulated using COMSOL Multiphysics simulation software. The magnetic fields are used to generate the graphs shown in Fig. \ref{experimentalSetup}. A linearization of the magnetic force field is used to simulate the trajectories of particles using a finite-difference approximation method simulator developed in-house. The simplified characterization of the magnetic force through linearization parameters are also useful in optimizing magnetic localization using the theoretical methods introduced here.

\end{methods}

\section{Supplementary Discussion}
\subsection{Pseudo-stability} \label{pseudo}

It is important to distinguish prior pseudo-localization methods and the method presented here. Although a magnetic energetic minimum cannot be achieved in 3D, pseudo-localization relies on restricting one dimension of particle movement do allow for an energy well in the remaining two dimensions. Particles will be attracted and aggregate to this well position in 2D. However, this aggregation method does not extend to 3D as particles would disperse through the third dimension. Our method focuses particles in two dimensions \textit{without} dispersing them in the third dimension. This enables true 3D aggregation.
\section{Supplementary Equations} \label{seq}
\subsection{Magnetic Field}
\hfill \break
Magnetostatic equations can be derived from Maxwell's equations,

\begin{equation}
\nabla\cdot{\bf D} = \rho_c
\end{equation}
\begin{equation}
\nabla\cdot{\bf B} = 0
\end{equation}
\begin{equation}
\nabla\times{\bf E} = - {\frac{\partial{\bf B}}{\partial t}} 
\end{equation}
\begin{equation}
\nabla\times{\bf H} = {\bf J} + {\frac{\partial{\bf D}}{\partial t}} 
\end{equation}

\noindent where $\mathbf{D}$ is the displacement field, $\bf{E}$ is the electric field, $\bf{B}$ is the magnetic flux density, $\bf{H}$ is the magnetic field, $\bf{J}$ is the current density, and $\rho_c$ is the charge density.

\noindent In magnetostatics, these equations reduce to,
\begin{equation}
\nabla\cdot{\bf B} = 0
\end{equation}
\begin{equation}
\nabla\times{\bf H} = {\bf J}
\end{equation}
The magnetic flux density is related to the magnetic field by
\begin{equation}\label{blinear}
\mathbf{B}\equiv \mu_{0}(\mathbf{H} + \mathbf{M})
\end{equation}
where $\mu_0$ is the permeability of free space and $\mathbf{M}$ is the magnetic dipole moment per unit volume. For a linear media, $\mathbf{M}$ has a linear relationship on the magnetic field and is given by
\begin{equation}
\mathbf{M}=\chi_{m} \mathbf{H}
\end{equation}
where $\chi_{m}$ is the media susceptibility. Thus, (\ref{blinear}) can be rewritten in a common form using the definition $\mu \equiv \mu_0(1+\chi_m)$ as
\begin{equation}
\mathbf{B}=\mu \mathbf{H}.
\end{equation}

\subsection{Magnetic Force on a Dipole Moment}
The potential energy $U$ of a magnetic dipole moment $\mathbf{m}$ is given by \cite{griffiths2005introduction}
\begin{equation}
\mathbf{U}=-\mathbf{m}\cdot \mathbf{B}_{ext}.
\end{equation}
Note that $\mathbf{B}_{ext}$ is the magnetic flux density in the absence of the magnetic dipole moment. The force on the magnetic dipole moment can be calculated from the potential energy by
\begin{equation} \label{force_dipole}
\mathbf{F}=-\nabla \mathbf{U}=\nabla(\mathbf{m}\cdot \mathbf{B}_{ext}).
\end{equation}
Because $\nabla \times \mathbf{B}_{ext}=0$, this can alternatively be written as 
\begin{equation} \label{force}
\mathbf{F} = (\mathbf{m}\cdot\nabla)\mathbf{B}_{ext}
\end{equation}
where $\mathbf{m}\cdot\nabla$ is the convective operator.


\subsection{Magnetic Force on a Paramagnetic Particle}
The magnetization of a particle in a homogeneous magnetizing field is affected by the demagnetization factor (\textit{N}) and is given by \cite{smistrup2005magnetic}
\begin{equation} \label{demag}
\mathbf{M} = \frac{\chi_i}{1+N\chi_i} \mathbf{H} .
\end{equation}
where $\chi_i$ is the intrinsic magnetic susceptibility of the particle.  For a spherical particle, $N = \frac{1}{3}$. The magnetic dipole moment for a spherical particle of volume $V$ is calculated from (\ref{demag}) and the identity $\mathbf{m}=\mathbf{M}V$ and is given by
\begin{equation}\label{moment}
\mathbf{m}=\chi V\mathbf{H},\quad \chi = \frac{3\chi_i}{3+\chi_i}
\end{equation}
The last step is to calculate the external magnetic flux density of the environment ($\mathbf{B}_{ext}$).
For most environments, like water, the permeability is approximately equal to the magnetic permeability constant because of a lack of magnetizable substances besides the particles. Therefore, 
\begin{equation}
\mathbf{B}_{ext} \approx \mu_{0}\mathbf{H}.
\end{equation}
Using this result as well as (\ref{force_dipole}) and (\ref{moment}), we find that the force on a spherical paramagnetic particle is given by
\begin{equation} \label{force}
\mathbf{F} = \frac{V\mu_0\chi}{2}\nabla\lVert \mathbf{H} \rVert ^2
\end{equation}

\subsection{Earnshaw's Theorem}
\hfill \break
Samuel Earnshaw proved that point charges can not attain a stability point in a static electric field. A similar proof states that a paramagnetic particle cannot achieve a stability point in a static magnetic field. For a slowly moving magnetic particle in a static magnetic field, the force is conservative because it depends only on the position of the particle and possibly its orientation (magnetic dipole). 
For a paramagnetic particle in a static magnetic field, the total potential energy $\textit{U}$ can be calculated from (\ref{force_dipole},\ref{force}) and is given by 
\begin{equation}
U= -k|\mathbf{H}|^{2} = -k(H_{x}^{2}+H_{y}^{2}+H_{z}^{2})
\end{equation}
where $k$ is a positive quantity equal to
\begin{equation} 
k = \frac{V\mu_0 \chi}{2}
\end{equation}
Any local minimum can be found for positive values of the Laplacian of the potential energy. The Laplacian of the potential energy is given by
\begin{equation}
\begin{aligned}
\nabla^{2} U &=  -2k(|\nabla H_{x}|^{2}+|\nabla H_{y}|^{2}+|\nabla H_{z}|^{2}\\
&+H_{x}\nabla^{2}H_x+H_{y}\nabla^{2}H_y+H_{z}\nabla^{2}H_z)\\
&=-2k\left(|\nabla H_{x}|^{2}+|\nabla H_{y}|^{2}+|\nabla H_{z}|^{2}\right)\\
& = -2k \left\| \nabla \mathbf{H}\right\|^2_{F}
\end{aligned}
\end{equation}
since $\nabla \cdot H = 0$, $ \nabla \times H= \mathbf{0}$ and $\nabla^2 H= \nabla\left(\nabla \cdot H\right) - \nabla \times \left(\nabla \times H \right)$. Thus, we have that
\begin{equation}
\nabla^{2} U \leq 0 
\end{equation}

\noindent The result is that there exist no local or global energetic minimums (or potential energy wells). However, the underlying assumption of Earnshaw's theorem is a static magnetic field. Although at any instant in time the magnetic force field is unstable, a collection of varying magnetic fields could create a stable system.

\subsection{Rotational Magnetic Force Stability in 2D} 
\hfill \break
\noindent Let's recall the definition of the magnetic potential 
\begin{equation}
U = -k|\mathbf{H}|^{2} = -k(H_{x}^{2}+H_{y}^{2}+H_{z}^{2})
\end{equation}
where $k$ is a positive quantity equal to
\begin{equation} 
k = \frac{V\mu_0\chi}{2}.
\end{equation}
and assume that it has an equilibrium at the origin $0$ (i.e. $\nabla U(0) =0$. To simplify notation, we will express $U$ as
\begin{align}
U = -kU_H,\quad U_{H} = \left| \mathbf{H}\right|^2.
\end{align}
Assume we have a spherical particle with density $\rho$, radius $r$ and subject to Stokes's drag in a viscous fluid with dynamic viscosity $\eta$. 
Then, if we rotate the potential $U(x)$ around its equilibrium at the frequency $\omega$, we obtain the particle dynamics
\begin{align}
\label{eq:absODE0}
\ddot{x}(t) + \frac{b}{m}\dot{x}(t) = \frac{k}{m}R(\omega t)\nabla U_{H}\left(R^\mathrm{T}\left(\omega t\right)
{x(t)}\right)
\end{align}
where $x(t) \in \mathbb{R}^2$ and $R\left(\omega t\right)$ denotes the rotation matrix function
\begin{align}
\label{eq:Rmat}
	R\left(\omega t\right) = \begin{bmatrix} \cos(\omega t)& -\sin(\omega t)\\\sin(\omega t) & \cos(\omega t) \end{bmatrix}.
\end{align}
and $b$, $m$ are computed as
\begin{align}
b = 6\pi \eta r,\quad m = V\rho
\end{align}
We will define the following constants
\begin{align}
\epsilon = \frac{b}{2m} = \frac{6 \pi \eta r}{8/3 \pi \rho r^3} = \frac{9 \eta}{4dr^2}, \quad \alpha = \frac{k}{m} = \frac{\mu_0 \chi}{2 \rho}
\end{align} 
to rewrite \eqref{eq:absODE0} as
\begin{align}
\label{eq:odeepsal}
\ddot{x}(t) + 2\epsilon\dot{x}(t) = \alpha R(\omega t)\nabla U_{H}\left(R^\mathrm{T}\left(\omega t\right)
{x(t)}\right).
\end{align}
In the following section, we will derive conditions on $\omega$, $\epsilon$, $\alpha$ and the curvature of the saddle equilibrium of $U_H$ that causes exponential stability in the 2D case. 
Furthermore, we will show that we can utilize this 2D saddle stabilization to accomplish control of particles in 3D. We will show that if we switch between two different 2D saddle stabilization rotations at the right interval, we can achieve net exponential stability in 3D.
For the purpose of the later stability analysis, we will consider only the linearized equations of \eqref{eq:odeepsal}: 
\begin{align}\label{eq:linODE}
\ddot{x}(t) + 2\epsilon\dot{x}(t) = \alpha R(\omega t)H_0R^\mathrm{T}\left(\omega t\right)
{x(t)}
\end{align}
where $H_0\in\mathbb{R}^{2\times 2}$ is the Hessian of $U_H$ evaluated at the origin.
Under the reasonable assumption that $U_H$ is a smooth function at the origin, the Hessian $H_0$ is symmetric and is diagonalizable with respect to an orthogonal basis at any point $x_0$, including the origin. Since we are rotating $U_H$, the particular orientation doesn't matter to the analysis, so w.l.o.g, we can assume that the Hessian $H_0$ is diagonalizable with respect to the coordinate axis. Therefore w.l.o.g., we will parametrize $H_0$ as the diagonal matrix $D(\mu,\delta)$ with the entries $c_\mu-c_\delta$ and $c_\mu +c_\delta$ on its diagonal:
\begin{align}\label{eq:defH0}
H_0 = \left.\begin{bmatrix} \frac{\partial^2 U_H}{\partial x^2_1} & \frac{\partial^2 U_H}{\partial x_2\partial x_1}\\ \frac{\partial^2 U_H}{\partial x_1\partial x_2} &\frac{\partial^2 U_H}{\partial x^2_2}\end{bmatrix}\right|_{0}= D(c_\mu,c_\delta):=  \begin{bmatrix} c_\mu-c_\delta & 0\\0 &c_\mu+c_\delta \end{bmatrix}.
\end{align}
With this particular parametrization of $H_0$ we can give an geometric interpretation to the quantities $c_\mu$ and $c_\delta$. Therefore, $2c_\mu$ is the divergence of $\nabla U_H$ or equivalently the Laplacian $\nabla^2 U_H$ at the origin
or the mean curvature of $U_H$ at the origin. A more negative $c_\mu$ implies a more "bowl" shape of the potential. A more positive $c_\mu$ implies a more "inverted or upside-down bowl" shape of the potential. On the other hand, $4c_\delta$ is difference between the maximal and minimal curvature of $U_H(x)$ at the origin and can be interpreted as the skewness or asymmetry of the curvature of $U_H(x)$. Thus, a larger $c_\delta$ implies a steeper saddle. We will therefore refer to $c_\mu$ and $c_\delta$ as the \textit{mean} and \textit{skewness} of $U_H$. These quantities can be computed in a coordinate independent way as
\begin{align}
c_{\mu} &= \frac{1}{2}\left.\nabla^2 U_H\right|_0\\
c_{\delta} &= \sqrt{\frac{1}{4}\left( \left.\nabla^2 U_H\right|_0\right)^2 - \mathrm{det}\left(\left. \mathcal{H}\left( U_H \right)\right|_0\right)}
\end{align}
where $\mathcal{H}$ denotes the Hessian operator.
\noindent To remove the time-varying character of the linear system \eqref{eq:linODE}, we will transform the dynamics \eqref{eq:linODE} with respect to the rotating coordinate system $z(t) = R^\mathrm{T}(\omega t)x(t)$ in which the potential force field becomes static. Substituting $x(t) = R(\omega t)z(t)$ and derivatives of same expression into equation \eqref{eq:linODE}, we obtain 
\begin{align}
\label{eq:relODE} &\ddot{z} + \left(2\omega S +2\epsilon I_2\right)\dot{z}+\left(2\omega\epsilon S- \omega^2I_2\right) z -\alpha D(c_{\mu},c_{\delta})z = 0.
\end{align}
Here we used that $R(\omega t)$ is unitary and the result of the calculation
\begin{align}\label{eq:RdotR}
\dot{R}\left(\omega t\right)R^\mathrm{T}\left(\omega t\right)=R^\mathrm{T}\left(\omega t\right)\dot{R}\left(\omega t\right) = \omega \begin{bmatrix}0&-1\\1 &0 \end{bmatrix} =:\omega S.
\end{align}

\noindent The advantage of representing the dynamics of the particle in its relative coordinates $z(t)$ is that the dynamics are time-invariant. Furthermore, $2\omega S \dot{z}$ and $-\omega^2 z$ are the well-known \textit{coriolis} and \textit{centripetal} force.
We will now perform yet another transformation to represent the dynamics with fewer dependent parameters. 

\noindent For the following derivation, notice that $\epsilon^2+\alpha c_{\mu} > 0$ holds for our magnetic potential. 
Let us introduce the following transformation, by representing $z(t)$ as
\begin{align}\label{eq:z2q}
z(t) = e^{-\epsilon t}q\left(\bar{\epsilon} t\right)&\,\,\mathrm{where}\,\,\bar{\epsilon} := \sqrt{\epsilon^2+\alpha c_{\mu}}.
\end{align}
Notice that we can always find a unique $q(\bar{t})$ to represent $z(t)$ in the form \eqref{eq:z2q}. In fact, the transformation between $q(\bar{t})$ and $z(t)$ is well-defined since it is a bijection with the inverse transformation 
\begin{align}\label{eq:q2z}
q(\bar{t}) = e^{\left(\epsilon/\bar{\epsilon}\right)t }z\left(\bar{t}/\bar{\epsilon}\right).
\end{align}
This transformation from $z(t)$ to $q(\bar{t})$ can be seen as an attempt to separate out the damping effect in the system together with a time-scaling transformation $\bar{t} = \bar{\epsilon} t$. Using the new representation \eqref{eq:z2q} of the relative particle position $z(t)$, we can now derive the dynamics for the component $q(\bar{t})$, by writing $z(t)$ and its derivatives in terms of $q(\bar{t})$ and plugging them into the $z(t)$ dynamics \eqref{eq:relODE}. In doing so, we derive the expressions
\begin{align}\label{eq:zder2q}
\dot{z}\left(t\right) &= -\epsilon e^{-\epsilon t}q\left(|\bar{\epsilon}| t\right) + |\bar{\epsilon}| e^{-\epsilon t}q'\left(|\bar{\epsilon}| t\right)\\
\ddot{z}\left(t\right) &= \epsilon^2 e^{-\epsilon t}q\left(|\bar{\epsilon}| t\right) - 2\epsilon|\bar{\epsilon}|e^{-\epsilon t}q'\left(\bar{\epsilon }t\right)+|\bar{\epsilon}|^2 e^{-\epsilon t}q''\left(|\bar{\epsilon}| t\right)
\end{align}
and after the mentioned substitution and multiplying by $e^{\epsilon t}$, we obtain the equations for $q(\bar{t})$ as:
\begin{align}\label{eq:q1dyn}
\notag\epsilon^2 q- 2\epsilon|\bar{\epsilon}|q'+|\bar{\epsilon}|^2 q'' + \left(2\omega S +2\epsilon I_2\right)\left(-\epsilon q + |\bar{\epsilon}| q'\right) + \hdots \\
+\left(2\omega\epsilon S- \omega^2I_2\right) q - \alpha D\left(c_{\mu},c_{\delta} \right)q = 0
\end{align}
where we denote $q'$ as $dq/d\bar{t}$ to explicitly differentiate between the different time scales $t$ and $\bar{t}$. Taking advantage of the occurring cancellations, equation \eqref{eq:q1dyn} can be further simplified to
\begin{align}\label{eq:q2dyn}
|\bar{\epsilon}|^2 q'' + 2\omega |\bar{\epsilon}|Sq'-(\epsilon^2 + \omega^2)q - \alpha D\left(c_{\mu},c_{\delta} \right)q = 0
\end{align}
Recalling the definition of $D$ in \eqref{eq:defH0} and dividing equation \eqref{eq:q2dyn} by $|\bar{\epsilon}|^2$, we can rewrite \eqref{eq:q2dyn} as:
\begin{align}\label{eq:q3dyn}
 q'' + 2\frac{\omega}{|\bar{\epsilon}|}Sq'-\left(\frac{\omega}{|\bar{\epsilon}|}\right)^2q -D\left(1,\frac{\alpha c_{\delta}}{|\bar{\epsilon}|^2} \right)q = 0
\end{align}
Finally, recalling the definition of $\bar{\epsilon}$ and introducing the normalized variables
\begin{align}\label{eq:normvar}
\bar{\omega} = \frac{\omega}{\bar{\epsilon}} = \frac{\omega}{\sqrt{\epsilon^2 + \alpha c_{\mu}}} && \bar{\delta} = \frac{\alpha c_\delta}{\bar{\epsilon}^2} = \frac{ c_\delta}{\epsilon^2/\alpha+ c_{\mu}}
\end{align}
we can represent the dynamics of $q$ as 
\begin{align}\label{eq:qdynmain}
 q'' + 2\bar{\omega}Sq'-\bar{\omega}^2q -D\left(1,\bar{\delta} \right)q = 0.
\end{align}
 Using standard procedures, we obtain the equation of the characteristic polynomial of the q-dynamics to be
\begin{align}\label{eq:qdyneig}
P_q(\lambda) &= \lambda^4 + 2\lambda^2\left(\bar{\omega}^2-1\right) + \left(\bar{\omega}^2 + 1\right)^2 -\bar{\delta}^2.
\end{align}
Furthermore, computing the roots of \eqref{eq:qdyneig} gives us the eigenvalues of the q-dynamics as
\begin{align}\label{eq:qeigs}
\lambda_i = \pm\sqrt{1 -\bar{\omega}^2 \pm\sqrt{\bar{\delta}^2 - 4\bar{\omega}^2}}. 
\end{align}

\noindent Now comparing the dynamic equations \eqref{eq:relODE} and \eqref{eq:qdynmain}, to which we will refer to as \textit{z-dynamics} and \textit{q-dynamics}, we can see that the system representation in the q-dynamics \eqref{eq:qdynmain} has a simpler representation. Indeed, the q-dynamics describe a frictionless version of the z-dynamics with a $\alpha c_{\mu} =1$. Moreover, the transformation \eqref{eq:z2q} can be interpreted to separate out the effect that friction directly has on the z-dynamics. In other words, this also demonstrates that in order to investigate the behavior of system A with parameters ($\omega_A, \alpha_A, c_{\mu_A},c_{\delta_A}, \epsilon_A$), it is equivalent to analyze system B with parameters ($\omega_B = \bar{\omega}_A, \alpha_{B}=1, c_{\mu} = 1,c_{\delta_B} = \bar{\delta}_A, \epsilon_B = 0$). Hence, by transforming to the q-dynamics, we managed to reduce the number of dependent variables from originally 5 ($\omega, \alpha, c_\mu,c_\delta, \epsilon$), to only two essential normalized variables $\bar{\omega}$, $\bar{\delta}$. For the reasons mentioned in the previous discussion, we will further on refer to $\bar{\omega}$ as the \textit{effective frequency} and to $\bar{\delta}$ as the $\textit{effective skewness}$ of the potential. Furthermore, the transformation to the original coordinate system is
\begin{align}
x(t) =e^{-\epsilon t}R(\omega t)q\left(\bar{\epsilon}t\right)
\end{align}

\noindent Recall the relationship between $x$, $\dot{x}$ and $q$, $q'$:
\begin{align}
x(t) &= e^{-\epsilon t}R(\omega t)q(\bar{\epsilon} t)\\
\dot{x}(t) &= e^{-\epsilon t}\left( \dot{R}(\omega t)q(\bar{\epsilon} t)-\epsilon R(\omega t)q(\bar{\epsilon} t) + \bar{\epsilon} R(\omega t)q'(\bar{\epsilon} t)  \right)\\
q(\bar{\epsilon} t) &= e^{\epsilon t}R^T(\omega t)x(t)\\
q'(\bar{\epsilon} t) &= \frac{1}{\bar{\epsilon}}e^{\epsilon t}\left( \dot{R}^T(\omega t)x(t)+\epsilon R^T(\omega t)x(t) + R^T(\omega t)\dot{x}(t) \right)
\end{align}
and notice that we can obtain the following bounds:
\begin{lem}\label{lem:x2q}
\begin{align*}
\left\| x(t)\right\|_2 &\leq e^{-\epsilon t}\left\| q(\bar{\epsilon} t)\right\|_2\\
\left\| \dot{x}(t)\right\|_2 &\leq \bar{\epsilon}e^{-\epsilon t}\left(\sqrt{(\epsilon/\bar{\epsilon})^2 +\bar{\omega}^2}\left\| q(\bar{\epsilon} t)\right\|_2 + \left\| q'(\bar{\epsilon} t)\right\|_2\right)\\
\left\| \dot{x}(t)\right\|_2 &\geq \bar{\epsilon}e^{-\epsilon t}\left(\sqrt{(\epsilon/\bar{\epsilon})^2 +\bar{\omega}^2}\left\| q(\bar{\epsilon} t)\right\|_2 - \left\| q'(\bar{\epsilon} t)\right\|_2\right)
\end{align*}
\end{lem}
\begin{lem}\label{lem:q2x}
\begin{align*}
\left\| q(\bar{\epsilon}t)\right\|_2 &\leq e^{\epsilon t}\left\| x( t)\right\|_2\\
\left\| q'(\bar{\epsilon} t)\right\|_2 &\leq e^{\epsilon t}\left(\sqrt{(\epsilon/\bar{\epsilon})^2 +\bar{\omega}^2}\left\| x(t)\right\|_2 +\frac{1}{\bar{\epsilon}} \left\| \dot{x}(t)\right\|_2\right)\\
\left\| q'(\bar{\epsilon} t)\right\|_2 &\geq e^{\epsilon t}\left(\sqrt{(\epsilon/\bar{\epsilon})^2 +\bar{\omega}^2}\left\| x(t)\right\|_2 -\frac{1}{\bar{\epsilon}} \left\| \dot{x}(t)\right\|_2\right)
\end{align*}
\end{lem}
\begin{proof}
The first inequality comes directly from the relation $x(t) = e^{-\epsilon t}R(\omega t)q(\bar{\epsilon} t)$ and the fact that $R(\omega t)$ is unitary. For the other two inequalities, we have to simplify the term $\dot{x}^T(t)\dot{x}(t)$: 
\begin{align}\label{eq:xdot_2}
\notag\dot{x}^T(t)\dot{x}(t)&= e^{-2 \epsilon t}\left(\left(\omega^2 + \epsilon^2\right) \left\|q(\bar{\epsilon}t) \right\|^2_2 +\bar{\epsilon}^2 \left\|q'(\bar{\epsilon}t) \right\|^2_2 -2 \epsilon \bar{\epsilon}q^T(\bar{\epsilon}t)q'(\bar{\epsilon}t) + 2\omega \bar{\epsilon} q^T(\bar{\epsilon}t)Sq'(\bar{\epsilon}t)\right)\\
\notag&=e^{-2 \epsilon t}\left(\left(\omega^2 + \epsilon^2\right) \left\|q(\bar{\epsilon}t) \right\|^2_2 +\bar{\epsilon}^2 \left\|q'(\bar{\epsilon}t) \right\|^2_2 + 2\bar{\epsilon} q^T(\bar{\epsilon}t)\left( \omega S - \epsilon I\right)q'(\bar{\epsilon}t) \right)
\end{align}
To this end, we can use the fact 
\begin{align}
(\omega S - \epsilon I)^T(\omega S - \epsilon I) = (\omega^2 + \epsilon^2)I
\end{align}
and Cauchy-Schwarz to obtain the bound 
\begin{align}
\notag \left|q^T(\bar{\epsilon}t)\left( \omega S - \epsilon I\right)q'(\bar{\epsilon}t) \right|
\leq &\left\|q(\bar{\epsilon}t)\right\|_2 \left\| \left(\omega S - \epsilon I\right)q'(\bar{\epsilon}t)\right\|_2  =\sqrt{\omega^2 + \epsilon^2}\left\|q(\bar{\epsilon}t)\right\|_2 \left\|q'(\bar{\epsilon}t)\right\|_2.
\end{align}
This allows us to derive from \eqref{eq:xdot_2}, the following the upper bound
\begin{align}
\dot{x}^T(t)\dot{x}(t)\leq&e^{-2 \epsilon t}\left(\left(\omega^2 + \epsilon^2\right) \left\|q(\bar{\epsilon}t) \right\|^2_2 +\bar{\epsilon}^2 \left\|q'(\bar{\epsilon}t) \right\|^2_2 + 2\bar{\epsilon} \sqrt{\omega^2 + \epsilon^2}\left\|q(\bar{\epsilon}t)\right\|_2\left\|q'(\bar{\epsilon}t)\right\|_2 \right)\\
\label{eq:xdotupper_2}\leq&e^{-2 \epsilon t}\left(\sqrt{\omega^2 + \epsilon^2}\left\|q(\bar{\epsilon}t)\right\|_2 + \bar{\epsilon}\left\|q'(\bar{\epsilon}t)\right\|_2 \right)^2
\end{align}
and the lower bound
\begin{align}\label{eq:xdotlower_2}
\dot{x}^T(t)\dot{x}(t)
\geq&e^{-2 \epsilon t}\left(\left(\omega^2 + \epsilon^2\right) \left\|q(\bar{\epsilon}t) \right\|^2_2 +\bar{\epsilon}^2 \left\|q'(\bar{\epsilon}t) \right\|^2_2 - 2\bar{\epsilon} \sqrt{\omega^2 + \epsilon^2}\left\|q(\bar{\epsilon}t)\right\|_2\left\|q'(\bar{\epsilon}t)\right\|_2 \right)\\
\geq&e^{-2 \epsilon t}\left(\sqrt{\omega^2 + \epsilon^2}\left\|q(\bar{\epsilon}t)\right\|_2 - \bar{\epsilon}\left\|q'(\bar{\epsilon}t)\right\|_2 \right)^2 .
\end{align}
By taking the square root and factoring out $\bar{\epsilon}$ we obtain from \eqref{eq:xdotlower_2} and \eqref{eq:xdotupper_2} the desired bounds. The proof for the other pair of inequalities is omitted in the interest of brevity but follows using similar arguments. 
\end{proof}
\noindent Now if $Re(\lambda_i)<\lambda_q, \forall i$, where $\lambda_i$ is an eigenvalue of the q-dynamics, then we know from linear control theory that the growth of the state $q$ and $q'$ can be bounded by
\begin{align}\label{eq:expq}
\left\|\begin{bmatrix}q(\bar{\epsilon} t) \\ q'(\bar{\epsilon} t) \end{bmatrix}\right\|_2 < k_q e^{\lambda_q\bar{\epsilon} t} \left\|\begin{bmatrix}q(0) \\ q'(0) \end{bmatrix}\right\|_2,
\end{align}
where $k_q >0$ is a constant bounding the transient behavior. Combining this together with the above bounds, we can conclude that $Re(\lambda_i)<\frac{\epsilon}{\bar{\epsilon}}, \forall i$ is a sufficient and necessary condition for exponential convergence of $x(t)$, as we will summarize in the next lemma:
\begin{lem}\label{thm:expconv}
Let $\lambda_i$ be the eigenvalues of the q-dynamics, then the linearized dynamics $x(t)$ are exponentially stable, i.e.
\begin{align}
\left\|\begin{bmatrix}x( t) \\ \dot{x}( t) \end{bmatrix}\right\|_2 < k_x e^{-\lambda_x t} \left\|\begin{bmatrix}x(0) \\ \dot{x}(0) \end{bmatrix}\right\|_2 
\end{align}
for some $k_x >0$, $\lambda_x > 0$ if and only if $Re(\lambda_i)<\frac{\epsilon}{\bar{\epsilon}}, \forall i$.
\end{lem}

\begin{proof}
$\Rightarrow$: $Re(\lambda_i)<\frac{\epsilon}{\bar{\epsilon}}, \forall i$ implies
\begin{align}\label{eq:expq}
\left\|\begin{bmatrix}q(t) \\ \dot{q}( t) \end{bmatrix}\right\|_2=\left\|\begin{bmatrix}q(\bar{\epsilon} t) \\ q'(\bar{\epsilon} t) \end{bmatrix}\right\|_2= \left\|\begin{bmatrix}q(\bar{t} ) \\ q'(\bar{t} ) \end{bmatrix}\right\|_2 < k_q e^{\frac{\epsilon}{\bar{\epsilon}} \bar{t}} \left\|\begin{bmatrix}q(0) \\ \dot{q}(0) \end{bmatrix}\right\|_2 < k_q e^{\epsilon t} \left\|\begin{bmatrix}q(0) \\ \dot{q}(0) \end{bmatrix}\right\|_2
\end{align}
and with Lem.\eqref{lem:x2q} we obtain
\begin{align}\label{eq:x2q0}
\left\|\begin{bmatrix} x(t)\\ \dot{x}(t) \end{bmatrix}\right\|_2 &\leq c_1 k_q e^{-\lambda_x t}\left\|\begin{bmatrix} q(0)\\ q'(0) \end{bmatrix}\right\|_2
\end{align}
\noindent where $c_1 = \bar{\epsilon}\sqrt{(\epsilon/\bar{\epsilon})^2 +\bar{\omega}^2} + 1+\bar{\epsilon}$, $\lambda_x >0$ with use of the triangle-inequality. Similarly, by using the opposite bounds Lem.\eqref{lem:q2x}, we can derive
\begin{align}\label{eq:q02x0}
\left\|\begin{bmatrix} q(0)\\ q'(0) \end{bmatrix}\right\|_2 \leq \frac{1}{\bar{\epsilon}}c_1 \left\|\begin{bmatrix}x(0) \\ \dot{x}(0) \end{bmatrix}\right\|_2.
\end{align}
Furthermore, from \eqref{eq:x2q0} and \eqref{eq:q02x0} we obtain 
\begin{align}
\left\|\begin{bmatrix}x( t) \\ \dot{x}( t) \end{bmatrix}\right\|_2 < \frac{1}{\bar{\epsilon}}k_qc^2_1 e^{-\lambda_x t} \left\|\begin{bmatrix}x(0) \\ \dot{x}(0) \end{bmatrix}\right\|_2,
\end{align}
which proves exponential stability of $x(t)$ with $k_x = (1/\bar{\epsilon})k_qc^2_1$.\\

\noindent $\Leftarrow$: If $x(t)$ is exponentially stable, then 
\begin{align}
\left\|\begin{bmatrix}x(t) \\ \dot{x}(t) \end{bmatrix}\right\|_2 \leq k_x e^{-\lambda_x t}\left\|\begin{bmatrix}x(0) \\ \dot{x}(0) \end{bmatrix}\right\|_2
\end{align}
holds for some $\lambda_x>0$ and similarly to the previous proof, we can conclude from \eqref{lem:q2x} 
\begin{align*}
&&\left\|\begin{bmatrix} q(\bar{\epsilon} t)\\ q'(\bar{\epsilon} t) \end{bmatrix}\right\|_2 &\leq e^{\epsilon t} \frac{1}{\bar{\epsilon}}c_1 \left\|\begin{bmatrix}x(t) \\ \dot{x}(t) \end{bmatrix}\right\|_2 \leq \frac{1}{\bar{\epsilon}}c_1 k_x e^{(\epsilon-\lambda_x) t}\left\|\begin{bmatrix}x(0) \\ \dot{x}(0) \end{bmatrix}\right\|_2\\
\Rightarrow &&\left\|\begin{bmatrix} q(\bar{ t})\\ q'(\bar{ t}) \end{bmatrix}\right\|_2 &< \frac{1}{\bar{\epsilon}}c_1 k_x e^{\frac{\epsilon}{\bar{\epsilon}}  \bar{t}}\left\|\begin{bmatrix}x(0) \\ \dot{x}(0) \end{bmatrix}\right\|_2\\
\Rightarrow&& Re(\lambda_i) &< \frac{\epsilon}{\bar{\epsilon}}
\end{align*}
where $\lambda_i$ are the eigenvalues of the $q$-dynamics
\end{proof}

\subsection{Condition for Exponential Stability}\hfill \break
\noindent This condition for exponential stability in the inertial frame can be expressed by the following conditions on the q-dynamics parameters:

\begin{lem}\label{lem:asympt}
The condition $Re(\lambda_i) < \frac{\epsilon}{\bar{\epsilon}}$ is equivalent to the following statements:
\begin{enumerate}[(i)]
\item \label{eq:cond} $
\bar{\omega} > \sqrt{1 -(\epsilon/\bar{\epsilon})^2}$ and $
2\sqrt{1-(\epsilon/\bar{\epsilon})^2 }\sqrt{\bar{\omega}^2+(\epsilon/\bar{\epsilon})^2 }< \bar{\delta} <\sqrt{4\bar{\omega}^2 + \left(\bar{\omega}^2 -1 + (\epsilon/\bar{\epsilon})^2\right)^2}$
\item \label{eq:cond2}
$\bar{\delta}>2\sqrt{1-(\epsilon/\bar{\epsilon})^2}$ and 
 $\sqrt{\sqrt{\bar{\delta}^2+4 (\epsilon/\bar{\epsilon})^2}-1-(\epsilon/\bar{\epsilon})^2} < \bar{\omega} < \sqrt{\frac{\bar{\delta}^2}{4\left( 1-(\epsilon/\bar{\epsilon})^2\right)} - (\epsilon/\bar{\epsilon})^2}$
\end{enumerate}
\end{lem}
\begin{proof}
We will first prove the equivalence of \eqref{eq:cond} and afterwards derive \eqref{eq:cond2} from \eqref{eq:cond}.\\
Recall that the eigenvalues of the q-dynamics can be calculated as
\begin{align}\label{eq:lam_i}
\lambda_i = \pm\sqrt{1 -\bar{\omega}^2 \pm\sqrt{\bar{\delta}^2 - 4\bar{\omega}^2}}.
\end{align}
In the following arguments, we will reformulate $Re(\lambda_i) < \frac{\epsilon}{\bar{\epsilon}}$ by exploiting the symmetric constellation of the eigenvalues \eqref{eq:lam_i}. To this end, we will separately analyze the cases $\bar{\delta} > 2\bar{\omega}$ and $\bar{\delta} \leq 2\bar{\omega}$:

\noindent \textbf{Case} $\bar{\delta} > 2\bar{\omega}$:\\
If $\bar{\delta}^2 > 4\bar{\omega}^2$, then inspecting \eqref{eq:lam_i}, we see that the eigenvalues come in pairs that are purely real or purely imaginary. Independent from the specific constellation of the eigenvalues
\begin{align}
\max_i \mathrm{Re}(\lambda_i)  = \sqrt{1 -\bar{\omega}^2 +\sqrt{\bar{\delta}^2 - 4\bar{\omega}^2}}
\end{align}
and therefore $\max_{i} Re(\lambda_i) < \epsilon/\bar{\epsilon}$ is satisfied iff
\begin{align}
& &1 -\bar{\omega}^2 +\sqrt{\bar{\delta}^2 - 4\bar{\omega}^2} &< \left(\epsilon/\bar{\epsilon}\right)^2\\
&\Leftrightarrow & \sqrt{\bar{\delta}^2 - 4\bar{\omega}^2} &< \bar{\omega}^2 -1 + \left(\epsilon/\bar{\epsilon}\right)^2\\
\label{eq:stat1}&\Leftrightarrow& \bar{\delta}^2 - 4\bar{\omega}^2 &< \left(\bar{\omega}^2 -1 + \left(\epsilon/\bar{\epsilon}\right)^2\right)^2 \wedge \bar{\omega}^2 > 1-\left(\epsilon/\bar{\epsilon}\right)^2 
\end{align}
Overall, we showed that under the assumption $\bar{\delta} > 2\bar{\omega}$, the statement $\mathrm{Re}(\lambda_i)<\left(\epsilon/\bar{\epsilon}\right)$ is equivalent to:
\begin{align}\label{eq:case1}
\bar{\omega} > \sqrt{1-\left(\epsilon/\bar{\epsilon}\right)^2} \wedge \bar{\delta}<\sqrt{4\bar{\omega}^2 + \left(\bar{\omega}^2 -1 + \left(\epsilon/\bar{\epsilon}\right)^2\right)^2} 
\end{align}

\noindent \textbf{Case} $\bar{\delta} \leq 2\bar{\omega}$:\\
If this holds, notice from the symmetric constellation of the eigenvalues in \eqref{eq:lam_i}, that they have to take the form $\lambda_i = \pm a \pm bi$. Hence, the characteristic polynomial can be alternatively represented as
\begin{align}
&P_q(\lambda)\\
=&\left(\lambda-a+bi \right)\left(\lambda-a-bi \right)\left(\lambda+a-bi \right)\left(\lambda+a+bi \right)\\
= &\lambda^4+2\left(b^2-a^2\right)\lambda^2 + \left(a^2+b^2\right)^2.
\end{align}
By comparing coefficients with the original parametrization of the characteristic polynomial
\begin{align}
P_q(\lambda) = \lambda^4+2\lambda^2\left(\bar{\omega}^2-1 \right) + \left(\bar{\omega}^2 + 1 \right)^2- \bar{\delta}^2.
\end{align}
we get the correspondence
\begin{align}
\left(a^2+b^2\right)^2 &= \left(\bar{\omega}^2 + 1 \right)^2- \bar{\delta}^2 \\
a^2-b^2 &= 1-\bar{\omega}^2.
\end{align}
or equivalently
\begin{align}
\left(a^2+b^2\right)^2 &= \left(1 - \bar{\omega}^2 \right)^2 + 4\bar{\omega}^2 - \bar{\delta}^2 \\
a^2-b^2 &= 1-\bar{\omega}^2.
\end{align}
From this, we can conclude
\begin{align}
1-\bar{\omega}^2 + \sqrt{\left(1 - \bar{\omega}^2 \right)^2 + 4\bar{\omega}^2 - \bar{\delta}^2} = 2a^2
\end{align}
and due to $\mathrm{Re}(\lambda_i) < c \Leftrightarrow -c<a<c \Leftrightarrow a^2 < c^2$, we have 
\begin{align}
&&\mathrm{Re}(\lambda_i) &< \left(\epsilon/\bar{\epsilon} \right)\\
&\Leftrightarrow & 1-\bar{\omega}^2 + \sqrt{\left(1 - \bar{\omega}^2 \right)^2 + 4\bar{\omega}^2 - \bar{\delta}^2} &< 2\left(\epsilon/\bar{\epsilon} \right)^2
\end{align}
We can simplify this last condition as
\begin{align}
& & 1-\bar{\omega}^2 + \sqrt{\left(1 - \bar{\omega}^2 \right)^2 + 4\bar{\omega}^2 - \bar{\delta}^2} &< 2\left(\epsilon/\bar{\epsilon} \right)^2 \nonumber\\
\label{eq:cd2}&\Leftrightarrow & \left(1 - \bar{\omega}^2 \right)^2 + 4\bar{\omega}^2 - \bar{\delta}^2 <\left(\bar{\omega}^2-1+2\left(\epsilon/\bar{\epsilon} \right)^2\right)^2 &\wedge \bar{\omega}^2> 1 -2\left(\epsilon/\bar{\epsilon} \right)^2
\end{align}
Noting that 
\begin{align}
\left(1 - \bar{\omega}^2 \right)^2 + 4\bar{\omega}^2 - \left(\bar{\omega}^2-1+2\left(\epsilon/\bar{\epsilon} \right)^2\right)^2
=&4\bar{\omega}^2-4\left(\epsilon/\bar{\epsilon} \right)^4-4\left(\epsilon/\bar{\epsilon} \right)^2\bar{\omega}^2+4\left(\epsilon/\bar{\epsilon} \right)^2\\
=&4\left(1-\left(\epsilon/\bar{\epsilon} \right)^2 \right)\left(\bar{\omega}^2+\left(\epsilon/\bar{\epsilon} \right)^2 \right)
\end{align}
we can further simplify the condition \eqref{eq:cd2} to
\begin{align}\label{eq:case2_1}
&\bar{\omega}^2> 1 -2\left(\epsilon/\bar{\epsilon} \right)^2 \wedge  4\left(1-\left(\epsilon/\bar{\epsilon} \right)^2 \right)\left(\bar{\omega}^2+\left(\epsilon/\bar{\epsilon} \right)^2 \right) <\bar{\delta}^2.
\end{align}
Finally, we notice that under our case assumption $\bar{\delta} \leq 2\bar{\omega}$ the condition \eqref{eq:case2_1} has no solution for $\bar{\omega}$ taking values in the range $1 -\left(\epsilon/\bar{\epsilon} \right)^2 \geq \bar{\omega}^2> 1 -2\left(\epsilon/\bar{\epsilon} \right)^2$. Therefore, for $\bar{\delta} \leq 2\bar{\omega}$, we can reduce \eqref{eq:case2_1} equivalently to
\begin{align}\label{eq:case2}
&\bar{\omega}>\sqrt{ 1 -\left(\epsilon/\bar{\epsilon} \right)^2} \wedge  2\sqrt{1-\left(\epsilon/\bar{\epsilon} \right)^2 }\sqrt{\bar{\omega}^2+\left(\epsilon/\bar{\epsilon} \right)^2 } <\bar{\delta}.
\end{align}
Hence, we showed that under the assumption $\bar{\delta} \leq 2\bar{\omega}$, the condition $\mathrm{Re}(\lambda_i)<\left(\epsilon/\bar{\epsilon}\right)$ is equivalent to \eqref{eq:case2}.\\

\noindent Summarizing our analysis of both cases \eqref{eq:case1} and \eqref{eq:case2}, we showed that  $\mathrm{Re}(\lambda_i) < \left(\epsilon/\bar{\epsilon} \right)$ holds if and only if the condition
\begin{align}\label{eq:part1}
\bar{\omega} > \sqrt{1-\left(\epsilon/\bar{\epsilon}\right)^2} \wedge 2\bar{\omega}<\bar{\delta}<\sqrt{4\bar{\omega}^2 + \left(\bar{\omega}^2 -1 + \left(\epsilon/\bar{\epsilon}\right)^2\right)^2} 
\end{align}
or the condition
\begin{align}\label{eq:part2}
&\bar{\omega}>\sqrt{ 1 -\left(\epsilon/\bar{\epsilon} \right)^2} \wedge  2\sqrt{1-\left(\epsilon/\bar{\epsilon} \right)^2 }\sqrt{\bar{\omega}^2+\left(\epsilon/\bar{\epsilon} \right)^2 } <\bar{\delta} \leq 2\bar{\omega}
\end{align}
is satisfied. We can form the union or logical disjunction of  \eqref{eq:part2} and \eqref{eq:part1} to obtain the equivalence
\begin{align} 
\label{eq:delbound}&&\mathrm{Re}(\lambda_i)& <\left(\epsilon/\bar{\epsilon}\right) \\
 &\Leftrightarrow&\bar{\omega} >\sqrt{ 1 -\left(\epsilon/\bar{\epsilon} \right)^2} \quad\wedge \quad& 2\sqrt{1-\left(\epsilon/\bar{\epsilon} \right)^2} \sqrt{\bar{\omega}^2+\left(\epsilon/\bar{\epsilon} \right)^2 }< \bar{\delta} <\sqrt{4\bar{\omega}^2 + \left(\bar{\omega}^2 -1 + \left(\epsilon/\bar{\epsilon} \right)^2\right)^2} \nonumber
\end{align}
With \eqref{eq:delbound} we proved the \eqref{eq:cond} part of the Lemma. We will now reformulate \eqref{eq:delbound} to obtain \eqref{eq:cond2} of the Lemma.\\
Reformulating the upper-bound on $\bar{\delta}$ from \eqref{eq:delbound}, we have
\begin{align}
&& \bar{\delta} &<\sqrt{4\bar{\omega}^2 + \left(\bar{\omega}^2 -1 + \left(\epsilon/\bar{\epsilon} \right)^2\right)^2}\nonumber\\
&\Leftrightarrow& \bar{\delta}^2 &<4\bar{\omega}^2 + \left(\bar{\omega}^2 -1 + \left(\epsilon/\bar{\epsilon} \right)^2\right)^2 \\
& \Leftrightarrow& \bar{\delta}^2 &<\left(\bar{\omega}^2 +1 + \left(\epsilon/\bar{\epsilon} \right)^2\right)^2 - 4 \left(\epsilon/\bar{\epsilon} \right)^2\nonumber\\
& \Leftrightarrow& \bar{\delta}^2 + 4 \left(\epsilon/\bar{\epsilon} \right)^2&<\left(\bar{\omega}^2 +1 + \left(\epsilon/\bar{\epsilon} \right)^2\right)^2 \nonumber\\
\label{eq:wlp}& \Leftrightarrow& \sqrt{\bar{\delta}^2 + 4 \left(\epsilon/\bar{\epsilon} \right)^2} -1 -\left(\epsilon/\bar{\epsilon} \right)^2  &< \bar{\omega}^2.
\end{align} 
By considering $0<\left(\epsilon/\bar{\epsilon} \right)^2<1$, we can rewrite the lower-bound on $\bar{\delta}$ from \eqref{eq:delbound} as 
\begin{align}
&& 2\sqrt{1-\left(\epsilon/\bar{\epsilon} \right)^2} \sqrt{\bar{\omega}^2+\left(\epsilon/\bar{\epsilon} \right)^2 }&< \bar{\delta}\nonumber\\
&\Leftrightarrow & 4\left(1-\left(\epsilon/\bar{\epsilon} \right)^2 \right)\left(\bar{\omega}^2+\left(\epsilon/\bar{\epsilon} \right)^2 \right) &< \bar{\delta}^2\nonumber\\
\label{eq:wup}&\Leftrightarrow& \bar{\omega}^2&< \frac{\bar{\delta}^2}{4\left(1-\left(\epsilon/\bar{\epsilon} \right)^2 \right)}-\left(\epsilon/\bar{\epsilon} \right)^2.
\end{align}
From \eqref{eq:wup}, \eqref{eq:wlp} and \eqref{eq:delbound} we obtain that $\mathrm{Re}(\lambda_i) < \left(\epsilon/\bar{\epsilon} \right)$ is equivalent to 
\begin{align}\label{eq:iipart0}
\sqrt{\bar{\delta}^2 + 4 \left(\epsilon/\bar{\epsilon} \right)^2} -1 -\left(\epsilon/\bar{\epsilon} \right)^2<\bar{\omega}^2< \frac{\bar{\delta}^2}{4\left(1-\left(\epsilon/\bar{\epsilon} \right)^2 \right)}-\left(\epsilon/\bar{\epsilon} \right)^2 \quad \wedge \quad \bar{\omega}^2>1-\left(\epsilon/\bar{\epsilon} \right)^2. 
\end{align}
Notice that \eqref{eq:iipart0} implies
\begin{align}
\frac{\bar{\delta}^2}{4\left(1-\left(\epsilon/\bar{\epsilon} \right)^2 \right)}-\left(\epsilon/\bar{\epsilon} \right)^2>\bar{\omega}^2>1-\left(\epsilon/\bar{\epsilon} \right)^2
\end{align}
and is only feasible if $\bar{\delta}>2\sqrt{1-\left(\epsilon/\bar{\epsilon} \right)^2}$
and that 
\begin{align}\sqrt{\bar{\delta}^2 + 4 \left(\epsilon/\bar{\epsilon} \right)^2} -1 -\left(\epsilon/\bar{\epsilon} \right)^2<\bar{\omega}^2 \quad \wedge \quad \bar{\delta}>2\sqrt{1-\left(\epsilon/\bar{\epsilon} \right)^2}
\end{align} 
implies $\bar{\omega}^2>1-\left(\epsilon/\bar{\epsilon} \right)^2$. Therefore, \eqref{eq:iipart0} is equivalent to 
 \begin{align}
 \sqrt{\sqrt{\bar{\delta}^2+4 \left(\epsilon/\bar{\epsilon} \right)^2}-1-\left(\epsilon/\bar{\epsilon} \right)^2} < \bar{\omega} < \sqrt{\frac{\bar{\delta}^2}{4\left( 1-\left(\epsilon/\bar{\epsilon} \right)^2\right)} - \left(\epsilon/\bar{\epsilon} \right)^2} \quad \wedge \quad \bar{\delta}>2\sqrt{1-\left(\epsilon/\bar{\epsilon} \right)^2}
 \end{align}
which proves part \eqref{eq:cond2} of the Lemma.
\end{proof}

\noindent Inspecting Lemma \eqref{lem:asympt}, we can also see that the following lower-bounds on the effective frequency and skewness
\begin{align}
\bar{\delta} &>2\sqrt{1-\left(\epsilon/\bar{\epsilon} \right)^2}\\
\bar{\omega}&>\sqrt{1-\left(\epsilon/\bar{\epsilon} \right)^2}
\end{align}
are necessary conditions for exponential stability. 

\subsection{Condition for Exponential Stability in Original Parameters} \hfill \break
Recall our definitions of the effective frequency and skewness
\begin{align}\label{eq:normvar2}
\bar{\omega} = \frac{\omega}{\bar{\epsilon}} = \frac{\omega}{\sqrt{\epsilon^2 + \alpha c_{\mu}}} && \bar{\delta} = \frac{\alpha c_\delta}{\bar{\epsilon}^2} = \frac{ c_\delta}{\epsilon^2/\alpha+ c_{\mu}} && \bar{\epsilon} = \sqrt{\epsilon^2 + \alpha c_{\mu}}
\end{align}
where $\alpha$ and $\epsilon$ were defined as
\begin{align*}
\epsilon = \frac{b}{2m} = \frac{6 \pi \eta r}{8/3 \pi \rho r^3} = \frac{9 \eta}{4\rho r^2}, \quad \alpha = \frac{k}{m} = \frac{\mu_0 \chi}{2 \rho}.
\end{align*}
To have a representation of the exponential stability criterion in Lemma \eqref{lem:asympt} in terms of the original parameters, we can substitute the above expressions into the inequalities in Lemma \eqref{lem:asympt} and obtain
\begin{align}\label{eq:condpar1}
\omega > \sqrt{\alpha c_{\mu}} \quad \wedge \quad 2\sqrt{\frac{\omega^2}{\alpha c_{\mu}}+\frac{\epsilon^2}{\alpha c_{\mu}}} < \frac{c_{\delta}}{c_{\mu}}<\sqrt{4\frac{\omega^2}{\alpha c_{\mu}}\left(\frac{\epsilon^2}{\alpha c_{\mu}} +1\right)+\left(\frac{\omega^2}{\alpha c_{\mu}} +1\right)^2}
\end{align}
and 
\begin{align}\label{eq:condpar2}
\frac{c_{\delta}}{c_{\mu}} > 2\sqrt{\frac{\epsilon^2}{\alpha c_{\mu}}+1} \quad \wedge \quad  \sqrt{\alpha c_{\mu}}\sqrt{\sqrt{\frac{c^2_{\delta}}{c^2_{\mu}}+4\frac{\epsilon^2}{\alpha c_{\mu}}\left(\frac{\epsilon^2}{\alpha c_{\mu}}+1 \right)}-2\frac{\epsilon^2}{\alpha c_{\mu}}-1} < \omega < \sqrt{\alpha c_{\mu}}\sqrt{\frac{c^2_{\delta}}{4c^2_{\mu}} - \frac{\epsilon^2}{\alpha c_{\mu}}}
\end{align}
as an alternative but equivalent criterion for exponential stability. Moreover, we obtain the following lower-bounds on the frequency $\omega$ and the ratio $c_\delta/c_{\mu}$ as necessary conditions for exponential stability: 
\begin{align*}
\omega > \sqrt{\alpha c_{\mu}} && \frac{c_{\delta}}{c_{\mu}} > 2\sqrt{\frac{\epsilon^2}{\alpha c_{\mu}}+1}
\end{align*}

\subsection{Rotational Magnetic Force Stability in 3D}\label{sec:3Dcase}\hfill \break
In the previous section we discussed how we can control particles in two dimensions through saddle stabilization. This section will investigate how these ideas can be useful for the general three dimensional case.

\noindent We will show that it is possible to achieve asymptotic convergence in all three coordinates, by alternating between two saddle stabilization regimes. The approach can be modeled with the following switching dynamics:
\begin{align}
\notag &(\textbf{Regime }S_{12})&\quad \forall(t\text{ mod }2T)\in[0,T]: m\ddot{x}_{12}(t) + b\dot{x}_{12}(t) &= R(\omega t)D(\mu,\delta)R^\mathrm{T}\left(\omega t\right)
{x_{12}(t)}\\
\label{reg:S12} &&m\ddot{x}_3(t) + b\dot{x}_3(t) &= \mu_u x_3(t)\\
&(\textbf{Regime }S_{13})& \quad \forall(t\text{ mod }2T)\in[T,2T]:
 \notag m\ddot{x}_{13}(t) + b\dot{x}_{13}(t) &= R(\omega t)D(\mu,\delta)R^\mathrm{T}\left(\omega t\right)
{x_{13}(t)}\\
&&\label{reg:S13} m\ddot{x}_2(t) + b\dot{x}_2(t) &= \mu_u x_2(t).
\end{align}

\noindent In each of the regimes $S_{12}$ and $S_{23}$, we are controlling the position of the particle in the $x_1, x_2$ and $x_2, x_3$ plane by saddle stabilization for $T$ seconds respectively. On the other hand, the particle coordinates $x_3$ and $x_2$ are left uncontrolled and evolve with unstable dynamics  in the corresponding regime. As previously discussed, in each regime saddle stabilization is achieved by parameterizing $c_\mu$, $c_\delta$ and $\omega$ according to the conditions \eqref{eq:condpar1} and \eqref{eq:condpar2} to achieve exponential stability in the inertial frame. Despite allowing the particle to escape in one direction during each regime, we will show that if the aggregation in the two controlled dimensions is faster than the instability in the third uncontrolled direction, this approach can overall stabilize the particle in three dimensions. In particular, it is then possible to choose a large enough $T$, such that we can compensate for the drift in the corresponding uncontrolled third direction and achieve overall net exponential stability. 

\begin{thm}[Existence of Stabilizing $T$ under Unstable Drifts]\label{thm:altSad}
Consider the switched dynamics, described by the cases \eqref{reg:S12} and \eqref{reg:S13} and assume that $c_\mu$, $c_\delta$ and $\omega$ have been chosen according to the conditions \eqref{eq:condpar1} or \eqref{eq:condpar2}. Hence, without switching, the regimes $S_{12}$ and $S_{13}$ would decay exponentially w.r.t. to the bound:
\begin{align*}
&\left\|
\begin{bmatrix} x_{s}(t)\\ \dot{x}_{s}(t) \end{bmatrix}
\right\|_{2} \leq k_s e^{-\lambda_s t}\left\|\begin{bmatrix} x_{s}(0)\\ \dot{x}_{s}(0) \end{bmatrix}\right\|_2.
\end{align*}
while the unstable drift dynamics satisfy
\begin{align*}
&\left\|
\begin{bmatrix} x_{u}(t)\\ \dot{x}_{u}(t) \end{bmatrix}
\right\|_{2} \leq k_u e^{\lambda_u t}\left\|\begin{bmatrix} x_{u}(0)\\ \dot{x}_{u}(0) \end{bmatrix}\right\|_2.
\end{align*}
with the indeces $s$, $u$ representing the corresponding stable and unstable coordinates of the regimes. Then, the following holds:\\
Assume $\lambda_s > \lambda_u$ and $0< \bar{\lambda} < (\lambda_s-\lambda_u)/2$, then 
\begin{align}
\forall T \geq \max\left\{\frac{\log(2k_uk_s)}{\lambda_s-\lambda_u-2\bar{\lambda}},\frac{\log(2k^2_s)}{2(\lambda_s-\bar{\lambda})} \right\}
\end{align}
the overall switched dynamics \eqref{reg:S12}, \eqref{reg:S13} are exponentially stable in $x(t)$ with at least the decay rate $\bar{\lambda}$.
\end{thm}
\begin{proof}
Since $c_\mu$, $c_\delta$ and $\omega$ have been chosen according to \eqref{eq:condpar1}, \eqref{eq:condpar2}, due to Lemma \eqref{lem:asympt} the dynamics \eqref{reg:S12} and \eqref{reg:S13} are exponentially stable w.r.t. to $x_{12}(t)$ and $x_{13}(t)$. Hence $\forall t \in [2nT,2nT+T]$ holds
\begin{align}\label{eq:contbound1}
&\left\|
\begin{bmatrix} x_{12}(t-2nT)\\ \dot{x}_{12}(t-2nT) \end{bmatrix}
\right\|_{2} \leq k_s e^{-\lambda_s (t-2nT)}\left\|\begin{bmatrix} x_{12}(2nT)\\ \dot{x}_{12}(2nT) \end{bmatrix}\right\|_2
\end{align}
and $\forall t \in [2nT+T,2nT+2T]$ holds
\begin{align}\label{eq:contbound2}
&\left\|
\begin{bmatrix} x_{13}(t-(2n+1)T)\\ \dot{x}_{13}(t-(2n+1)T) \end{bmatrix}
\right\|_{2} \leq k_s e^{-\lambda_s (t-(2n+1)T)}\left\|\begin{bmatrix} x_{13}(2n+1)T)\\ \dot{x}_{13}(2n+1)T \end{bmatrix}\right\|_2
\end{align}
for some $k_s$, $\lambda_s > 0$ and any $n\in \mathbb{N}$. Furthermore, the drift of the uncontrolled coordinates is bounded $\forall t \in [2nT,2nT+T]$ as
\begin{align}\label{eq:driftx3}
\left\|
\begin{bmatrix} x_{3}(t-2nT)\\ \dot{x}_{3}(t-2nT) \end{bmatrix}
\right\|_{2} 
\leq & k_u e^{\lambda_u (t-2nT)} \left\|\begin{bmatrix} x_{3}(2nT)\\ \dot{x}_{3}(2nT)
 \end{bmatrix}\right\|_2
\end{align}
and 
$\forall t \in [(2n+1)T,(2n+1)T+T]$ as
\begin{align}
\label{eq:driftx2} \left\|
\begin{bmatrix} x_{2}(t-(2n+1)T)\\ \dot{x}_{2}(t-(2n+1)T) \end{bmatrix}
\right\|_{2}
\leq  k_u e^{\lambda_u (t-(2n+1)T)}\left\|\begin{bmatrix} x_{2}((2n+1)T)\\ \dot{x}_{2}((2n+1)T) \end{bmatrix}\right\|_2
\end{align}
With these bounds we can analyze how $\left\|\left[x^T(2nT),\dot{x}^T(2nT) \right]^T \right\|_2$ evolves w.r.t. to the discrete time steps $\left\{0,2T,4T,\dots \right\}$. Towards this end, notice that we obtain
\begin{align}
&\left\|\begin{bmatrix} x(2T)\\ \dot{x}(2T) 
\end{bmatrix}\right\|_2\\ 
\leq & \left\|\begin{bmatrix} x_{13}(2T)\\ \dot{x}_{13}(2T) 
\end{bmatrix}\right\|_2 + \left\|\begin{bmatrix} x_{2}(2T)\\ \dot{x}_{2}(2T) 
\end{bmatrix}\right\|_2
\leq k_s e^{-\lambda_s T} \left\|\begin{bmatrix} x_{13}(T)\\ \dot{x}_{13}(T) 
\end{bmatrix}\right\|_2 + k_u e^{\lambda_u T}\left\|\begin{bmatrix} x_{2}(T)\\ \dot{x}_{2}(T) 
\end{bmatrix}\right\|_2 \\
\leq & \sqrt{2} \max\left\{k_u e^{\lambda_u T},k_s e^{-\lambda_s T}  \right\}\left\|\begin{bmatrix} x_{12}(T)\\ \dot{x}_{12}(T) 
\end{bmatrix}\right\|_2 + k_s e^{-\lambda_s T} \left\|\begin{bmatrix} x_{3}(T)\\ \dot{x}_{3}(T) 
\end{bmatrix}\right\|_2 \\
\notag\leq &  \sqrt{2}\max\left\{k_uk_se^{(\lambda_u-\lambda_s) T},k^2_s e^{-2\lambda_s T}  \right\}\left\|\begin{bmatrix} x_{12}(0)\\ \dot{x}_{12}(0) 
\end{bmatrix}\right\|_2 + k_s k_u e^{(\lambda_u-\lambda_s) T}\left\|\begin{bmatrix} x_{3}(0)\\ \dot{x}_{3}(0) 
\end{bmatrix}\right\|_2\\
\label{eq:maxineq}\leq & \max\left\{2k_uk_se^{(\lambda_u-\lambda_s) T},2k^2_s e^{-2\lambda_s T}  \right\}\left\|\begin{bmatrix} x(0)\\ \dot{x}(0) 
\end{bmatrix}\right\|_2
\end{align}
where we made multiple use of the norm-inequality
\begin{align} \label{eq:normineq}
\frac{1}{\sqrt{2}}\left(\left\|x\right\|_2 + \left\|y\right\|_2\right) \leq \left\|\begin{bmatrix}
x\\y\end{bmatrix}\right\|_2 \leq \left\|x\right\|_2 + \left\|y\right\|_2.
\end{align}
Recall that we assumed $\lambda_s > \lambda_u$ and picked $0< \bar{\lambda} < (\lambda_s-\lambda_u)/2$ with $T$ satisfying 
\begin{align}\label{eq:Tcond}
 T \geq \max\left\{\frac{\log(2k_uk_s)}{\lambda_s-\lambda_u-2\bar{\lambda}},\frac{\log(2k^2_s)}{2(\lambda_s-\bar{\lambda})} \right\}.
\end{align}
Using these assumptions we obtain from \eqref{eq:Tcond} that the following holds:
\begin{align}\label{eq:Tcond1}
  T &\geq \frac{\log(2k_uk_s)}{\lambda_s-\lambda_u-2\bar{\lambda}} 
 \Leftrightarrow T\left(\lambda_s-\lambda_u-2\bar{\lambda}\right) \geq \log(2k_uk_s)  \Leftrightarrow e^{-2\bar{\lambda}T} \geq 2k_uk_se^{\left(\lambda_u-\lambda_s \right)T}
\end{align}
\begin{align}\label{eq:Tcond2}
  T &\geq \frac{\log(2k^2_s)}{2\left(\lambda_s-\bar{\lambda}\right)} 
 \Leftrightarrow 2T\left(\lambda_s-\bar{\lambda}\right) \geq \log(2k^2_s)  \Leftrightarrow e^{-2\bar{\lambda}T} \geq 2k^2_se^{-2\lambda_sT}.
\end{align}
Therefore, from \eqref{eq:maxineq} we can conclude
\begin{align*}
\left\|\begin{bmatrix} x\left(2T\right)\\ \dot{x}\left(2T\right) 
\end{bmatrix}\right\|_2
\leq e^{-2\bar{\lambda}T}\left\|\begin{bmatrix} x(0)\\ \dot{x}(0) 
\end{bmatrix}\right\|_2
\end{align*}
and by induction we get
\begin{align}
\left\|\begin{bmatrix} x\left(2nT\right)\\ \dot{x}\left(2nT\right) 
\end{bmatrix}\right\|_2\leq  e^{-2\bar{\lambda}nT}\left\|\begin{bmatrix} x(0)\\ \dot{x}(0) 
\end{bmatrix}\right\|_2
\end{align}
which shows us that we can guarantee $\lim \limits_{n\rightarrow \infty}\left\|\left[x^T(2nT),\dot{x}^T(2nT) \right]^T \right\|_2 =0$.
Moreover, from \eqref{eq:contbound1}, \eqref{eq:contbound2}, \eqref{eq:driftx2}, \eqref{eq:driftx3} and noticing that $\left \lfloor \frac{t}{T} \right \rfloor T \geq t-T$ we can conclude:
\begin{align*}
&\left\|\begin{bmatrix} x(t) \\ \dot{x}(t) \end{bmatrix} \right\|_2\\ 
\leq &\max \left\{k_s, k_ue^{\lambda_u T}\right\} \left\|\begin{bmatrix} x\left(\left \lfloor \frac{t}{T} \right \rfloor T \right) \\ \dot{x}\left(\left \lfloor \frac{t}{T} \right \rfloor T \right) \end{bmatrix} \right\|_2
\leq \max \left\{k_s, k_ue^{\lambda_u T}\right\}^2 \left\|\begin{bmatrix} x\left(\left \lfloor \frac{t}{2T} \right \rfloor 2T \right) \\ \dot{x}\left(\left \lfloor \frac{t}{2T} \right \rfloor 2T \right) \end{bmatrix} \right\|_2\\
\leq &\max \left\{k_s, k_ue^{\lambda_u T}\right\}^2  e^{-\bar{\lambda}\left \lfloor \frac{t}{2T} \right \rfloor 2T}\left\|\begin{bmatrix} x(0)\\ \dot{x}(0) 
\end{bmatrix}\right\|_2
\leq \max \left\{k_s, k_ue^{\lambda_u T}\right\}^2  e^{-\bar{\lambda}(t-2T)}\left\|\begin{bmatrix} x(0)\\ \dot{x}(0) 
\end{bmatrix}\right\|_2\\
\leq &e^{2\bar{\lambda}T} \max \left\{k_s, k_ue^{\lambda_u T}\right\}^2 e^{-\bar{\lambda}t}\left\|\begin{bmatrix} x(0)\\ \dot{x}(0) 
\end{bmatrix}\right\|_2
\end{align*}
which proves exponential stability of the switched system with decay rate $\bar{\lambda}$ and concludes the proof. Notice that the corresponding transient constant is bounded by 
\begin{align}
\bar{k} \leq e^{2\bar{\lambda}T} \max \left\{k_s, k_ue^{\lambda_u T}\right\}^2
\end{align}
\end{proof}

\section{References}

\bibliography{conbib}


\begin{addendum}
 \item 
The authors would like to thank Michelle Wang for her indefatigable and unflappable efforts in helping to establish an experimental demonstration of aggregation. Alex White provided us with creative debugging and passionate spirits in our $11^th$ hour. The authors would also like to thank Brian Hong for his efforts in perusing and proofreading this manuscript. Last but not least, the authors would like to thank Reza Fatemi for extremely helpful technical discussions.
 \item[Competing Interests] The authors declare that they have no
competing financial interests.
 \item[Correspondence] Correspondence and requests for materials
should be addressed to Ali Hajimiri~(email: hajimiri@caltech.edu).
\end{addendum}


\end{document}